\documentclass{article}
\usepackage{spconf,amsmath,graphicx}
\usepackage{siunitx}
\usepackage{xcolor}
\usepackage{Here}
\usepackage{cite}

\title{Breaking Imaging Limits using Dithering Masks \\in 0.35 terahertz Single-Pixel Imaging}
\name{S. Augustin$^{1,2}$, T. Szollmann$^{3}$, P. Jung$^{3}$, H.-W. H\"{u}bers$^{1,2}$\thanks{Thanks to the German Research Foundation (DFG) for funding the work presented here in the scope of the priority program SPP1798 Compressed Sensing in Information Processing (CoSIP) ─ grants HU848/7 and JU2795/3.}}
\address{\small $^1$Humboldt Universit\"{a}t zu Berlin, Department of Physics, 10099 Berlin, Germany. \\
\small$^2$German Aerospace Center (DLR), Institute of Optical Sensor Systems, 12489 Berlin, Germany. \\
\small$^3$Technische Universit\"{a}t Berlin, Department of Electrical Engineering and Computer Science, 10587 Berlin, Germany.}
\begin{document}
%
\maketitle
\begin{abstract}
In this contribution, we explore the imaging limits of terahertz single-pixel cameras and present a way to improve the achievable signal-to-noise ratio and spatial resolution using a technique called \emph{dithering masks}. The proposed procedure is implemented on the hardware level using recent advances in the development of spatial light modulators for terahertz single-pixel cameras. Thereby, our approach offers new ways to overcome theoretical imaging limits of near wavelength resolution. We suggest that in using dithering masks the required phase retrieval driven recovery can be simplified to a robust non-negative linear model even for near wavelength terahertz single-pixel imaging.
\end{abstract}
\section{Motivation \& Introduction}\label{sec:motiv}
Single-pixel cameras (SPCs) \cite{Duarte.2008} have found widespread use in many areas both on the research level and also in the industrial domain. In particular, SPCs that work outside of the visible have quickened interests in many fields. The advantages of the single-pixel approach, which  are the possibility for signal processing on the hardware level (computational imaging) and the image acquisition with a stationary single detector pixel, are especially beneficial in fields where a single detector pixel is multiple times more expensive than a detector pixel in the VIS domain or where multiple pixel cameras are not readily available (due to various physical reasons). This is exactly true for the low terahertz (THz) domain and in particular for \SI{0.35}{\tera\hertz}.\\
\indent
\SI{0.35}{\tera\hertz} radiation is of special interest because it is electromagnetic radiation with a somewhat unique combination of advantageous properties. It is able to penetrate many optical obstructions like packaging material, clothing, insulation material, etc. This property is, of cause, similar to $X$-ray's. But on the other hand $X$-ray's are so-called ionizing radiation, which can damage living cells and in general are harmful to organic tissue. In contrast terahertz is non-ionizing radiation, which is harmless for living cells and also for organic tissue in general. Additionally, terahertz has the advantageous property to sufficiently interact with the material it penetrates, retaining the ability to make objects visible that cannot be visualized with other means. Applications that would make use of this unique combination of advantageous properties are for example a terahertz camera that is able to visualize the processes inside living cells or the behavior of electrons inside quantum electronic devices.\vspace{6pt}\\
{\bf Prior Work on THz single-pixel imaging:}
One of the first publications on the idea of THz single-pixel imaging is reported in \cite{Chan.2008}. Here, the spatially-structured masks were implemented as metallic filters that had to be changed by hand, due to the lack of suitable THz spatial light modulators (SLMs).  One of the first non-manual implementations is published in \cite{Busch.2012}. Subsequent publications by various groups adapted the single-pixel THz idea to improved setups with differing THz sources and also differing implementations of the THz-SLM \cite{Shrekenhamer.2013,Watts.2014,Xie.2013,Augustin.2015}. In recent years a group in England reported the application of the single-pixel idea for near field imaging with spatial resolution well-below the wavelength \cite{Stantchev.2016}. However, far field single-pixel THz imaging with well-below wavelength resolution has not been reported so far.\vspace{6pt}\\
{\bf Main Contribution:}
We establish the connection between single-pixel \SI{0.35}{\tera\hertz}	imaging and diffraction. 
The influence of diffraction prohibits the usual linear intensity model and leads therefore to a non-linear imaging model that is similar to the well-known phase-retrieval problem. Since such problems are numerically challenging it is necessary, for real-time camera operation, to have appropriate linear approximations. 
In this work we, therefore, propose the application of dithering masks. 
Commanding these masks to the SLM allows for a robust, linearized imaging operation of single-pixel THz cameras. Under these conditions recovery is achieved using an efficient, non-negative least squares approach (NNLS).\\ \\

\begin{figure}[H]
\centerline{\includegraphics[width=0.45\textwidth]{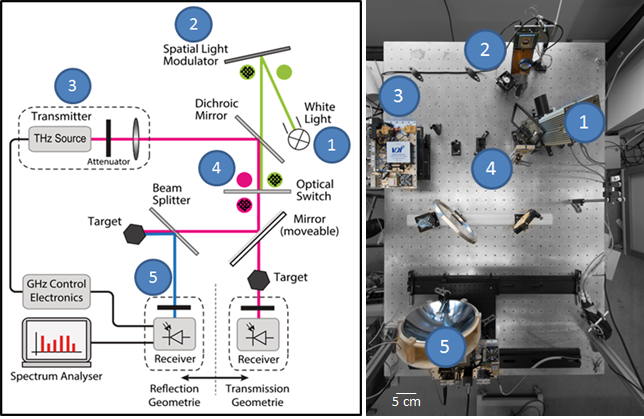}}
\caption{Representation of the 0.35 terahertz single-pixel camera used
for the experiments discussed here.\label{fig:cameraRepresentation}}
\end{figure}

\indent
The imaging modality for a \SI{0.35}{\tera\hertz} SPC that is of interest here is the use of the camera in a far-field imaging modality, where the scene is several centimeters away from the camera. This modality is especially appealing for the given envisioned application examples. To get an idea of the complexity and the distances involved, Fig. (\ref{fig:cameraRepresentation}) shows at the left-hand side a block diagram and in the right-hand side a representation of the \SI{0.35}{\tera\hertz} SPC used for the experiments discussed here. 
In this text we will explore the hardware signal processing ability and overall operation of a \SI{0.35}{\tera\hertz} SPC. To overcome the identified limitations a method based on dithering masks is presented, which will pave the way towards enabling the unique applications mentioned above. The text is thereby structured as follows. The next section introduces the terahertz single-pixel camera approach by discussing an imaging example and the respective measurement signals for standard \SI{0.35}{\tera\hertz} SPC operation. Section (\ref{sec:SPCimagingLIMITS}) then derives the limits of standard camera operation and supports them by both experimental and theoretical findings. Finally, the novel dither method is introduced.  
\section{Standard 0.35 terahertz single-pixel camera operation}
\begin{figure}
\centerline{\includegraphics[width=0.45\textwidth]{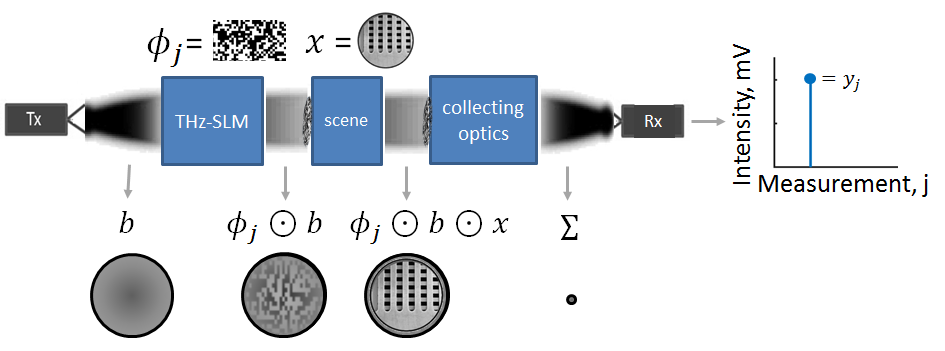}}
\caption{Visualization of the single-pixel camera measurement process according to the linear imaging model of \eqref{eqn:linearModel}.\label{fig:principle}}
\end{figure}

\begin{figure}[H]
\centerline{\includegraphics[width=0.45\textwidth]{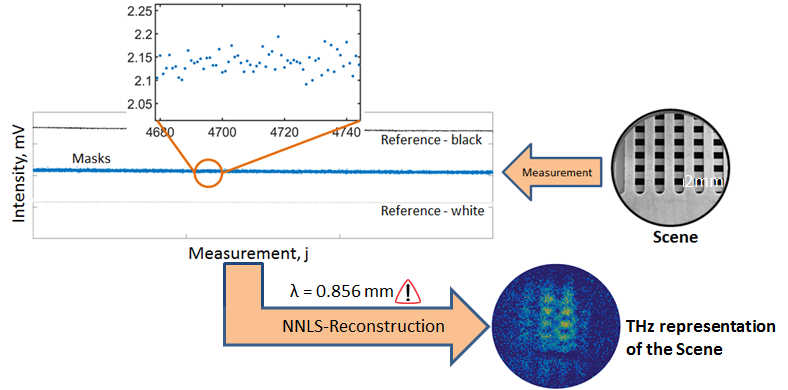}}
\caption{Example measurement and reconstruction of the measured values using the \SI{0.35}{\tera\hertz} SPC in standard measurement mode. As reconstruction algorithm non-negative least squares was used. Note, that the terahertz representation of the scene is inverted in regard to the photograph.\label{fig:figure 3}}
\end{figure}

A single-pixel camera is able to image a scene of interest with just a single detector pixel because the multiple acquisitions that are carried-out in parallel in a multi-pixel camera are carried-out sequentially in a single-pixel camera. The ability to substitute parallel image acquisition for sequential acquisition is possible due to the use of a spatial light modulator (SLM). The SLM is used to project patterns (masks) onto a scene of interest and the response from this projection
is measured by the single detector pixel.  The necessary number of sequential acquisitions depends (i) on the compressibility of the scene itself (like sparsity in a known basis) (ii) the properties of the masks and (iii) the method used for image reconstruction. The interconnection of the aforementioned aspects is non-linear and directly influences the performance of the imaging process in terms of spatial resolution, imaging speed, image quality, etc. Accordingly, the full model which is used to describe the imaging process is also non-linear. However, on a generic level, the standard operation of a \SI{0.35}{\tera\hertz} SPC can be described using a linearized imaging model as in \eqref{eqn:linearModel} and visualized in Fig. (\ref{fig:principle}).
\begin{equation}\label{eqn:linearModel}
  y = A\cdot x + n
\end{equation}
In \eqref{eqn:linearModel}, $y$ denotes the vector of measured values, $A$ is the measurement matrix, which contains in each row the effective masks used for the imaging process and $n$ models an additive noise component. $x$ is the (vectorized) object in the scene to be reconstructed from the measurements. A typical assumption in compressed sensing is that $x=\Psi\alpha$ where $\Psi$ is a known sparsifying transformation and $\alpha$ is a vector of transform coefficients that essentially concentrates only on few but unknown coordinates ($\alpha$ is sparse).\\
\indent
In this context, the inverse problem \eqref{eqn:linearModel} is meanwhile well-understood and, if possible, certain random matrices achieve optimal performance in terms of sampling rate and reconstruction accuracy. On the other hand, one of the main challenges in practical experiments is the optimization of the overall measurement setup in a way to come close to such a model. Often the linear model has (i) to be modified to account for other effects, like non-uniform illumination and (ii) is valid only in a sense of a certain approximation since, the detector may behave non-linearly and the other optical components have relevant influence.
Finally, (iii) diffraction will play a role.\\
\indent
In our case, we have to, additionally, account for the beam profile $b$, SLM effects and the detector measuring only integrated intensities.  As measurement
masks pseudo-random binary masks $\{\phi_j\}$ are commanded to the SLM and chosen here also due to their empirical robustness properties with respect to such model uncertainties (and their incoherence to a wide variety of transformations). Except of the effects due to (ii) and (iii), the imaging process is
further illustrated in Fig. (\ref{fig:principle}). The beam $b$ from the terahertz source (Tx) is spatially modulated by the $j$th mask $\phi_j$ commanded to the single-pixel cameras SLM. The spatially modulated beam $\phi_j\odot b$ is then used to probe the scene $x$ and the entire response $\phi_j\odot b\odot x$ from the scene is integrated into a single-measurement value $\langle \phi_j\odot b,x\rangle$ of the camera's single detector pixel. In short, every commanded mask leads to one measurement value. To image the entire scene a number of masks commensurate to the information content in the scene and the desired spatial resolution has to be measured. Following the imaging model \eqref{eqn:linearModel} an example measurement is shown in Fig. (\ref{fig:figure 3}).
On the left-hand side of the figure the measured mask characteristic is shown. The characteristic is comprised of three lines which show more or less variation depending on the respective commanded mask, model error and additive measurement noise. The top line shows the measurement response of commanded, spatially unstructured black masks. The lowest line results from commanded, spatially unstructured white masks. The middle line, which shows the largest variation, results form the spatially structured masks. Only this line carries spatial information about the scene. The information from the other two lines is simply used to adapt the terahertz camera to changing measurement conditions (measurement overhead). The information encoded in the variation of the middle line is used to reconstruct a terahertz representation of the measured scene. The reconstruction is shown in the bottom part of Fig. (\ref{fig:figure 3}).\\
\indent
The reconstruction shows that the periodic structure of the scene object was captured with a spatial resolution better than the wavelength. However, as is visible the spatial resolution is changing over the field-of-view of the camera and the reconstruction quality itself is also not homogeneous. The edges of some of the features in the image are not straight and are smeared out over several pixels. These artifacts are partially caused by deviations from the initial linear model \eqref{eqn:linearModel}. In the next section we will, therefore, present an imaging method which accounts for diffraction effects and still comes with the advantage of a linear imaging model. Starting from a full nonlinear imaging model that is related to the well-known phase-retrieval problem we will establish a modified
description that is applicable to single-pixel imaging with dithering masks.
\section{Imaging Limits of standard 0.35 terahertz single-pixel cameras}\label{sec:SPCimagingLIMITS}
\begin{figure}[H]
\centerline{\includegraphics[width=0.39\textwidth]{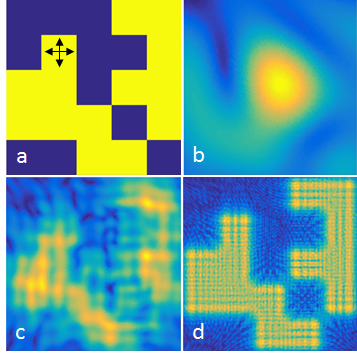}}
\caption{Simulation showing the distortions between the commanded masks (a) and the imaging masks for the 0.35 terahertz single-pixel camera considering mask block sizes of \SI{1}{\milli\meter} (b), \SI{3}{\milli\meter} (c) and \SI{10}{\milli\meter} (d). The arrows indicate a random mask block. All images b-d also show some undersampling artifacts of the simulation.\label{fig:figure4}}
\end{figure}

\begin{figure}[H]
\centerline{\includegraphics[width=0.415\textwidth]{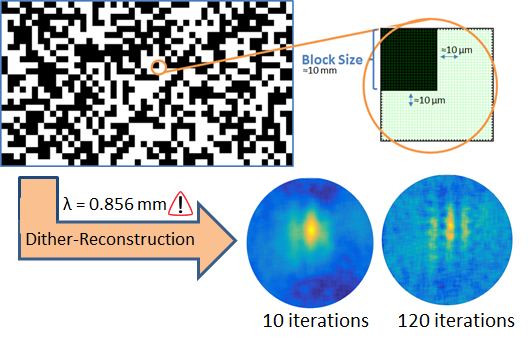}}
\caption{Illustration of the construction of a dithering mask set and a reconstruction using such a set (visualization of shifts not up to scale).\label{fig:dither}}
\end{figure}

As introduced in the last section, it is hypothesized that the limits of standard single-pixel camera operation stem from an inadequate imaging model. The linearized model Eqn. (\ref{eqn:linearModel}) is working adequately when the structures (block size) in the commanded masks are multiple times larger than the wavelength (\SI{0.35}{\tera\hertz} $\equiv$ \SI{0.85}{\milli\meter}). On the other hand the size of the mask structures are directly responsible for the spatial resolution that is achievable with a single-pixel camera. It means that the standard operation of a \SI{0.35}{\tera\hertz} SPCs is limited to spatial
resolution at the order of the wavelength. The cause of this limitation can be derived from the imaging model when the full imaging model is analyzed. In the following we will sketch particular steps towards a linear model and discuss some difficulties, which can arise for a practical device.\\
\indent
The starting point is the well-known diffraction integral \cite{Kirchhoff.1883}. For brevity of exposition we use generic two-dimensional coordinates $(u,v)$ without discretization.  Assume that the detector position is at $z=0$ on the optical axis and denote with $\mathcal{T}$ the target surface at position $z=d_{\mathcal{T}}$. Then, the $j$th measurement of the target $x=x(u,v)$ with an effective mask $\phi^{\text{THz}}_j=\phi^{\text{THz}}_j(u,v)$ -- which may differ
from commanded masks $\phi_j$ -- given (up to additive noise) as the magnitude of the following diffraction integral\footnote{We use  $a\sim b$ to indicate that, up to constants, $a$ is equal to $b$ (approximation).}.
\begin{equation}\label{eqn:fullModel}
  y_j \sim \left| \iint_{\mathcal{T}} \phi_j^{\text{THz}} \cdot x \frac{e^{ikr}}{r} \mathrm{d}u\mathrm{d}v\right|,
\end{equation}
with $r=\sqrt{u^2+v^2+d_{\mathcal{T}}^2}$  and wave number $k$.
With the far field assumption (by using
a lens), we assume that up to further scaling  factors, the measurements can
be written as:
\begin{equation}\label{eqn:fullModel2}
  y_j \sim \left|\iint_{\mathcal{T}} \phi_j^{\text{THz}} \cdot
    x\,\mathrm{d}u\mathrm{d}v\right
  |,
\end{equation}
However, the effective masks $\phi_j^{\text{THz}}$ are obtained by transferring the commanded, binary masks $\phi_j$ from the mask surface
$\mathcal{M}$ at $z=d_{\mathcal{M}}$ to target scene $\mathcal{T}$ via a diffraction integral operator:
\begin{equation}
  \phi_j^{\text{THz}}(u',v')\sim\iint_{\mathcal{M}}(\phi_j b + (1-\phi_j)a)\frac{e^{ikr'}}{r'}\mathrm{d}u\mathrm{d}v
  \label{eq:diffop}
\end{equation}
with $r'=\sqrt{(u'-u)^2+(v'-v)^2+(d_{\mathcal{M}}-d_{\mathcal{T}})^2}$. Recall that $b=b(u,v)\geq a(u,v)=a\geq0$ are the beam profile and SLM offset.  Obviously, \eqref{eq:diffop} is related to the Fourier transform for $d_{\mathcal{M}}-d_{\mathcal{T}}\rightarrow\infty$. Thus, an important design criterion is to command masks with sufficiently large block size such that for small $z$ the operation above is close to the identity (See Fig. (\ref{fig:figure4}) for a simulation of one mask with varying block size as evidence for this statement.).  
Then \eqref{eqn:fullModel} is approximately (up to constants and additive noise):
\begin{equation}
  y_j \approx \iint_{\mathcal{T}} (\phi_j b + (1-\phi_j)a) \cdot
    x\,\mathrm{d}u\mathrm{d}v
\end{equation}
In practice the imaging device needs to be calibrated as well but this subject is still an open research topic. After suitable discretization and for given non-negative $a$ and $b$, the problem can be described again with a measurement matrix $A$ as \eqref{eqn:linearModel}. Due to non-negativity and additional properties of $A$ (see here the $\mathcal{M}^+$-criterion in \cite{kueng:16:nnls}) the target can be reconstructed using a non-negative least-squares (NNLS) approach:
\begin{equation}
  \min_{x\geq 0}\lVert Ax-y\rVert_{\ell_2}
\end{equation}
This problem can be solved using, for example, active set algorithms. Meanwhile also other very efficient methods are available, see e.g. \cite{kiSrDh12}. In particular, our results are obtained using \textsc{Flexbox} \cite{dirks2015flexbox}. Note that, due to the non-negativity this approach still allows for subsampling and is related to $\ell_1$-self-regularization, see \cite{Slawski2013a}. Indeed, for random iid. binary matrices even optimal subsampling can be achieved
for sparse $x$ \cite{kueng:16:nnls} and similar conclusion can be drawn for certain other sparsity domains $\Psi$.\\
\indent Obviously, the full nonlinear diffraction model would describe the interdependence between the measurement values $y$ and the wavelength $\lambda$ of the radiation used in the single-pixel camera. But the current state-of-the-art approaches in phase retrieval still require numerically challenging algorithms and guarantees for randomized but practically relevant measurement models are rare. So far, there are no accurate algorithms which can solve \eqref{eqn:fullModel} in a quasi real-time fashion. Therefore the question arises, whether it is possible to still work with a sufficiently accurate linearized model \eqref{eqn:linearModel} and at the same time, produce spatial resolution better than the size of the wavelength.\vspace{6pt}\\
{\bf Dithering Masks explained:}
To overcome the  wavelength limitations in THz single-pixel imaging we propose the use of dithering masks for improved single-pixel camera operation. Dithering masks thereby means the use of masks with a large block size. These blocks are then shifted by random offsets in the order of the wavelength. Such a mask set should theoretically provide a large signal-to-noise ratio while keeping the high-resolution capabilities of a single-pixel setup. Additionally, such a mask set should not suffer from the mask degradation effects introduced above. The construction of a dithering masks set as well as a reconstruction using such a mask set are illustrated in Fig. (\ref{fig:dither}).
As can be seen the reconstruction with the dither mask set shows more homogeneous reconstruction results than the standard reconstruction of Fig. (\ref{fig:figure 3}). The edges of the reconstructed features are straight and not smeared out over several pixels, indicating a better spatial resolution. However it is noteworthy that the reconstruction success crucially depends on the chosen reconstruction parameter combination and the number of iterations of the reconstruction algorithm.\vspace{6pt}\\
{\bf{Acknowledgements:}}
The authors are very grateful to Hendrik Dirks for his support with
the software reconstruction of the presented data using
\cite{dirks2015flexbox}.
\label{sec:refs}
%

\bibliographystyle{IEEEbib}
\bibliography{ICASSP2018}

\end{document}